\documentclass{sf2a-conf2015}
\usepackage{graphicx}
\usepackage{hyperref}
\usepackage[]{natbib}  
\usepackage{epstopdf}

\def\BibTeX{{\rm B\kern-.05em{\sc i\kern-.025em b}\kern-.08em
    T\kern-.1667em\lower.7ex\hbox{E}\kern-.125emX}}
\bibpunct{(}{)}{;}{a}{}{,}  

\begin{document}

\TitreGlobal{SF2A 2015}


\title{Atomic data needs for the modelling of stellar spectra}

\runningtitle{Atomic data and stellar spectra}

\author{R.Monier$^{1,}$}\address{LESIA, UMR 8109, Observatoire de Paris Meudon, Place J.Janssen, Meudon, France}\address{Lagrange, UMR 7293, Universite de Nice Sophia, Nice, France}

\setcounter{page}{237}


\maketitle


\begin{abstract}
The current need for atomic data to model stellar spectra obtained in various wavelength ranges is described.
The level of completeness and accuracy of these data is discussed.
\end{abstract}



\section{Introduction}
High quality atomic data are necessary for modelling properly the spectra of stars of various spectral types.
The nature of these various data is reviewed here. I also address two crucial issues: i) the completeness of this data and ii) their accuracy.


\section{Nature des donn\'ees atomiques}

Let us recall that for an atmosphere composed of one layer only, the line profile of an absorption line, $I_{\lambda}$,  depends on wavelength as
$$ I_{\lambda} = I_{0} \exp (-\tau_{\lambda}) $$
where $I_{0}$  stands for the adjacent continuum and  $\tau_{\lambda}$ is the optical depth which can be expressed as:
  $$ d \tau_{\lambda} = - (\kappa_{\lambda} + l_{\lambda}) \rho dz $$
where $\kappa_{\lambda}$ and $l_{\lambda}$ are the coefficients for continuous and line absorption respectively.
The line absorption coefficient can be expressed as:
$$l_{\lambda} = \lambda_{ij}^{2} f_{if} (\frac{N_{i}}{N}) V $$
where $\lambda_{ij}$ is the wavelength between the lower level  (i) and the higher level (j), $f_{ij}$ the oscillator strength, $\frac{N_{i}}{N}$ 
the ratio of absorbers over the total number contributing to the line and $V$ is the normalised Voigt profile.

Under the simple assumption of Local Thermodynamical Equilibrium (LTE), the ratio $\frac{N_{i}}{N}$ is given by the Saha and Boltzmann equations  and the parameters of the transition: the energy $E_{ij}$, 
the ionization potential IP, the statistical weight $g$ and the partition function.
In the laboratory, the spectroscopist measures the intensity of the line and its position and shape from an emission spectrum.  
The laser and beam-foil techniques allow to measure the lifetimes of the atomic energy levels and the relative intensities of lines.
From these measurements, one can infer:
   \begin{itemize}
  \item[$-$] the difference in energy between the two levels $E_{ij}$
  \item[$-$] the ionization potential IP
  \item[$-$] the degeneracy of the level $g$
  \item[$-$] the transition probability $A_{ij}$
  \item[$-$] the angular momentum quantum number $J$
  \item[$-$] the branching fraction $BF_{ij} = \frac{I_{ij}}{\sum I_{j}} $ 
  \end{itemize}

\section{Spectroscopic diagnosis}

The  local thermodynamical equilibrium (LTE) is verified in a stellar atmosphere when collision processes dominate over radiative processes to
populate the levels.
If this assumption cannot hold, one should resort to Non Local Thermodynamic Equilibrium (NLTE) and solve the equation of statistical equilibrium:
$$ \frac{dn_{i}}{dt} = \sum_{i \neq j}^{N} n_{j} P_{ji} - n_{i} \sum_{i \neq j}^{N} P_{ij} = 0 $$
where $P_{ij}$ is the transition probability:
$$ P_{ij} = R_{ij} (J_{\nu_{0}}) + C_{ij}(T) $$
where $R_{ij}$ and $C_{ij}$ refer to the radiative and collicional transitions, $J_{\nu_{0}}$ the local intensity field and $T$ the local temperature.
Resolving this equation requires the knowledge of the collision coefficients $C_{ij}$ for the excitation and ionisation of various species of all abundant absorbers.
It also requires the knowledge of the photoionization cross sections.
As a rule, the occurence of several ionization stages (neutral or ionized) in a stellar spectrum depends on the effective temperature $T_{eff}$.
Typically one does not observe more than three ionization stages in a photosphere if the star does not have a chromosphere, nor a corona.

\section{Historical perspective of the production of atomic data }

The National Bureau of Standards (NBS) first produced massively atomic data. I refer the reader to NBS monograph 53 published in
1962 (\cite{Corliss1962}). It is the first compilation of measured wavelengths and oscillator strengths for 25000 spectral lines observed for 70 chemical elements in the spectral range from 
2000 \AA\ up to 9000 \AA. This data are affected by systematic errors but some of them are still used today.
In 1960, the Massachusets Institute for Technology (MIT) published the MIT wavelength tables which contain about 100000 lines between 2000 \AA\ and 10000 \AA.
On the other hand, quantum computations have produced massively new atomic data following the MCHF and MCDF formalisms
(Multi Configuration Hartree Fock ou Dirac-Fock,Froese Fischer and Taschiev, see http: physics.nist.gov/MCHF, updated Sep 2010)).
Similarly Robert Kurucz has produced theoretical calculations 
 (\cite{Kurucz11}) and so has the team of the Opacity Project (see http:cdsweb.u-strasbg.fr/topbase/op.html)

More recently, the Vienna Atomic Line Database (VALD) has incorporated the new results produced by Kurucz 
(\cite{Kupka2000}).
The National Institute of Standards has produced the NIST database of critically evaluated data on atomic energy levels, wavelengths and transition probabilities
and of recent bibliographic data (\cite{Kramida13}, see http:www.nist.gov). The Troisk database also collects new experimental data (\cite{Kramid10}).
The trienial reports of Commission 14 (Atomic and Molecular data) of the International Astronomical Union are also important data sources.
The DREAM (Database on Rare Earths at Mons University, see w3.umh.ac.be/~astro/dream.shtml, \cite{Biemont05}) provides experimental measurements for the neutral and ionized Lanthanides.
\\
For each of these databases, the question of the completeness and accuracy arise.  It all depends on the goal one pursues.
The completeness is crucial for opacity computations important for modelling the  internal structure and the atmospheric structure of stars.
It is necessary to include all strong and weak lines of all elements even if the wavelengths and oscillator strengths are affected by errors.
On the other hand, for abundance or velocity fields determinations, one needs high quality atomic data whose errors 
should be critically evaluated.


\section{Stellar spectra at various resolutions}

At low resolution (R $\simeq$ 1000), the observed spectra only allow to locate the strongest absorption and emission lines. They can be used to assign a spectral type to a star
(Morgan Keenan (MK) spectral tagging). If, in addition, this low resolution data could be calibrated into absolute fluxes, they can be used to test model atmosphere predictions provided the angular diameter of the star is known at various wavelengths.
\\
At intermediate resolution (R $\simeq$ a few $10^{3}$), it is in principle possible to resolve a large number of lines if the lines are not smeared out by rapid rotation. Abundances can be determined by modelling these lines provided they have accurate atomic parameters and they are unblended. The blending depends on the rotation rate and on the individual abundances as well which are a priori unknown.

At high resolution (typically an instrument like the \'echelle cross-disperser SOPHIE at Observatoire de Haute Provence, R= 75000, or HARPS at ESO, R=115000 ), 
one can hope to measure subtle effects of atomic processes on the line profiles  such as, for example, isotopic shifts (IS) or hyperfine structure (HFS) or magnetic broadening due to a magnetic field or asymetries in the line wings which are due to depth-dependent convective motions.


\section{Problems specific to the ultraviolet region (UV)}

The accuracy and completeness of atomic data are rather poor in the UV for elements heavier than the iron-peak elements (Z $\geq$ 28).
However, for a few chemical elements, the UV is the only domain where their lines can be found.
We can quote \cite{Wahlgren2011} synthesis of the Se II line  (Z=34)  for the Copernicus spectrum of 3 Cen A (B5 IIIp) 
which exemplifies well the problem of accuracy.
There are no usable Se II lines in the optical range for abundance determinations so this constitutes the first detection of selenium in a star.
The oscillator strength for this line stems from a computation but its error is unknown ($\log gf = -0.32$).
Adopting this value , the selenium abundance , [Se/H]  is found to be 4 times the solar value.
However, if one increases $\log gf$ by +0.30, [Se/H] becomes solar. As a consequence, it is not possible to determine accurately the selenium abundance.

\subsection{Lack of completeness in the UV}

\begin{figure}[ht!]
\centering
 \includegraphics[width=1.0\textwidth,clip]{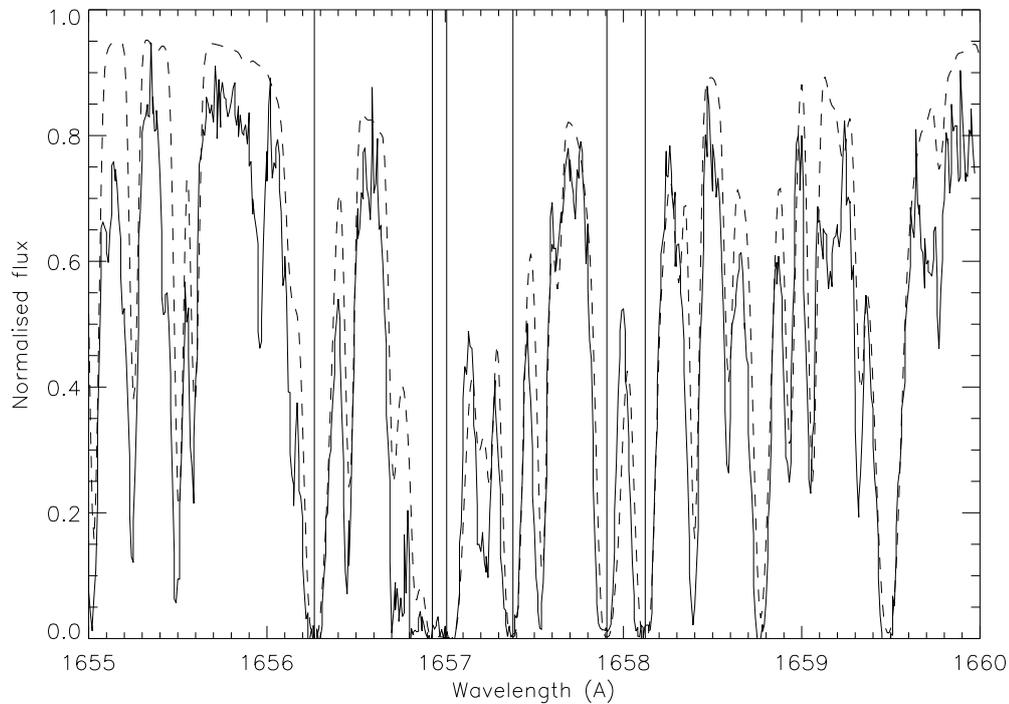}  \vskip -8.5cm   
 \caption{Synthesis of the C I UV Multiplet 1 for HD 72660 (observed: solid line, model: dashed line)}
   \label{fig1}
\end{figure}

The completeness of atomic data is also an issue in the UV.
We can take the example of the spectral synthesis of the hot Am star HD 72660 (\cite{Varenne99}) at wavelengths shorter than 
2000 \AA\ (the far UV). Over the entire far UV many observed lines in the STIS spectrum of HD 72660 do not have counterparts in the synthetic spectrum because their
wavelengths and or oscillator strengths are still unknown. So many observed lines cannot be identified.
Figure  ~\ref{fig1} displays the synthesis of the 1655 to 1660 \AA\ wavelength range of the STIS spectrum of HD 72660 which harbours the UV Multiplet 1 resonance lines of C I (indicated by vertical lines, the observed spectrum is in solid lines, the synthetic spectrum in dashed lines). The adopted carbon abundance is that derived from optical spectra by \cite{Varenne99}. The C I lines are properly reproduced, however a number of observed lines are not properly synthezised indicating probably that these lines are still missing in our current linelists.
Some of these lines could be Fe II lines measured by Fourier Transform Spectroscopy (FTS) whose oscillator strengths are still unknown.
This is due to limited staff resources: several ions should measured by FTS in the laboratory but there is not sufficient funding nor sufficient staff to carry out the experiments.
In order to observe at wavelengths shorter than 1400 \AA, specific optical materials are necessary 
 (MgF2 beamsplitter).
 Computed data have complemented laboratory data but their uncertainty is difficult to evaluate if an experimental measurement is not available.


\section{Atomic data in the optical}

Most of the high excitation lines of iron-peak elements  ($E_{low} \geq 60000 cm^{-1}$) do not have oscillator strengths.
These lines are formed deep in the atmosphere where LTE prevails and can be used to derive reliable abundances if they have reliable atomic data.
 
\begin{figure}[ht!]
 \centering
  \includegraphics[width=1.0\textwidth,clip]{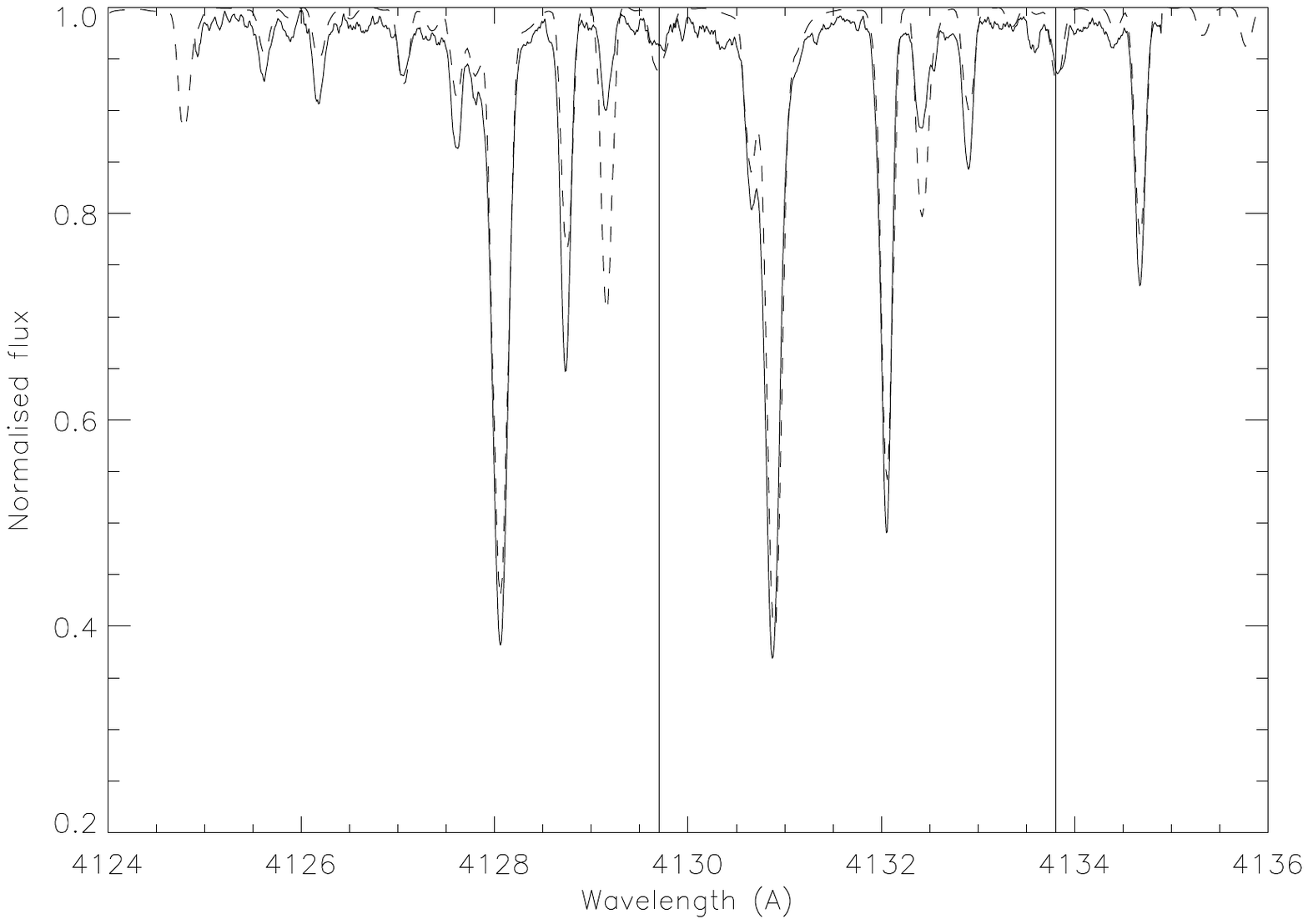}  \vskip -8.5cm   
   \caption{Detection of Ce II and Eu II lines in HD 72660  (observed: solid line, model: dashed line)}
   \label{fig2}
\end{figure}

The hot Am star HD 72660 (A1m) whose lines are very sharp ($v\sin i = 6 km.s^{-1}$) can be used to illustrate this problem.
The chemical composition of this star is fairly well known from the analysis of lines having accurate NIST data (\cite{Varenne99}). Using this composition as input to
the computation of a synthetic spectrum, one realises three types of problems: i) the oscillator strengths of many lines of the iron-peak elements
must be erroneous, ii) weak lines due to the Lanthanides (Ba, Nd, Ce, Eu,...) are poorly synthesized and iii) a number of observed lines remain unidentified.
Figure ~\ref{fig2} displays the synthesis of the HARPS spectrum of HD 72660 from 4124 \AA\ to 4136 \AA. The resonance line of Eu II at 4129.70 \AA (with its hyperfine structure of the various isotopes of Europium included) and  the Ce II line at 4133.80 \AA\ are clearly detected and are properly reproduced by mild overabundances of about 5 times solar for Europium and 20 times solar for Cerium. 


\subsection{The case of the low metallicity stars}

 Several rapid (r) and slow (s) neutron capture elements have been discovered in old stars of very low metallicities.
This discovery sets important constraints on the ancient chemical history of the Galaxy.
The first works emphasized the detection of light neutron capture elements (Sr, Y, Zr) and of Rare Earths (Ba, La and Eu).
For instance, \cite{Gilroy88} compared two metal poor giants of similar fundamental parameters:
  HD 122563  was found not to be enriched in neutron capture elements whereas HD 115444 turned out to be enriched in neutron capture elements.
  However, at the time of their analysis, \cite{Gilroy88} could only use a few transitions because few transition probabilities were known and their uncertainties were unknown. Actually, \cite{Gilroy88} used \cite{Corliss1962} data for the lines of Dy II and Gd II to derive the abundances of Dysprosium and Gadolinium.


The discovery of cool stars enriched in neutron capture elements has trigered laboratory work. As from 1973, Emile Bi\'emont and his collaborators in Belgium
have produced data for Y II, Zr II and Eu II.
Other groups in London, Lund and the University of Wisconsin have produced numerous transition probabilities for the Rare Earths from Baryum 
 to Hafnium (Z=72). The teams in London and Lund have focused on elements heavier than Thalium (Ta, Z=73) up to Uranium (U, Z=92).
These new data have allowed to greatly improve the abundance patterns of neutron capture elements for metal poor stars (\cite{Sneden14}).
Currently, the abundances of 37 neutron capture elements have been determined in CS 22892-052 (\cite{Sneden09}).


\section{Atomic data in the Infrared (IR)}

The infrared is an important wavelength range for cool stars as their spectral energy distribution reaches a maximum at 
$\lambda \geq 1 \mu m$. The infrared is also instrumental for the study of magnetic fields as the magnetic splitting is proportional to 
  $ \lambda^{2}$. It is also important  for the measurements of accurate radial velocities, to study molecular species as CO or SiO and atoms (C and N, neutron capture elements) which
  do not have many lines in the optical range.
 The IR is also useful to characterize exo-planets between 5 and 10 $\mu$m and circumstellar disks.
 Currently, only 118 oscillator strengths are available at wavelengths greater than 1 $\mu m$.
For practically all elements, laboratory spectra are missing for $\lambda \geq 1 \mu m$ precluding identifications.
Oscillator strengths are missing mostly because branching ratios $BF_{ij}$ are missing.



The branching ratios $Bf_{ij}$ are combined to the lifetimes of the levels $\tau_{j}$ in order to determine the transition probabilities $A_{ij}$
                          $$ A_{ji} =  {BF_{ji}  \over \tau_{j}} $$
			  
and the oscillator strengths $f_{ij}$ are infered by : 
$$ f_{ji} = 1.499 \times 10^{-8} {g_{j} \over g_{i}} \lambda^{2} A_{ji} $$

The accuracy on $f_{ij}$ thus depends on the accuracies on $\lambda$, $\tau$ et $BF_{ij}$.
The main difficulty is to estimate the uncertainty on the $BF_{ij}$.
For solar type stars, spectra like those obtained in the ACE (Advanced Chemistry Experiment) experiment give access to lines which are not acessible in the optical range.
The infrared will eventually allow studies of slow and rapid neutron capture elements in evolved stars and in Chemically Peculiar stars.



\section{Hyperfine structure and isotopic shifts}

The interaction of the magnetic moment of the nucleus , $\vec{\mu_{I}} = g {e \over 2 m_{p}} \vec{I}$ 
with the magnetic field, $\vec{B_{J}}$, generated by the orbital motion of the electrons modifies the fine structure.
The energy levels are displaced by the quantity: 
$$ \Delta E_{nhfs} = - \vec{\mu_{I}}.\vec{B_{J}} = {A \over \hbar^{2}} (\vec{I}.\vec{J})$$
The quantum number F verifies:
  $$\vec{F} = \vec{I} + \vec{J} $$
 and varies from  I+J to $|I-J|$).
 
 Each fine structure level is divided into 2I+1 sublevels if $I \leq J $ and into 2J+1 sublevels if $I \geq J$.
 The magnetic hyperfine splitting $\Delta E_{mhfs}$ verifies:
$$ \Delta E_{mhfs} =({A \over 2}) [F(F+1)-I(I+1)-J(J+1)] $$

In order for instance to properly model the Ni I line at 7414.5 \AA\ in the solar spectrum it is essential to include the hyperfine structure of the 5 isotopes of Nickel:
from $Ni^{58}$ up to $Ni^{64}$ (\cite{Sneden14}). The isotopic shifts for Nickel are close to 0.1 \AA\ and are more easily detectable in the infrared.


 

\section{Conclusion}

Accurate atomic data are urgently needed in the ultraviolet.  The ultraviolet spectra of many elements are still not properly understood.
The iron-peak elements have high priorities, in particular Fe II which is ubiquitous and contributes to blends at all wavelengths.
In the infrared very few transitions have oscillator strengths. Improving this situation will be essential to the future spatial missions in the IR (SOFIA and JWST)
or the VISIR spectrograph at VLT which will be working in the N and Q bands at R=150-30000.
For elements having several isotopes, it is highly desirable to include the hyperfine structure and the isotopic shifts for each isotope in order to properly model the observed line profiles at high resolution.


\begin{acknowledgements}
The author thanks Prof. Lydia Tchang-Brillet (Universit\'e Pierre et Marie Curie) for her advice at an early stage of this project.
\end{acknowledgements}


\bibliographystyle{aa}  
\bibliography{sf2a-template} 

\end{document}